\documentclass[useAMS,usenatbib]{mn2e}
\usepackage[dvips]{graphicx}
\usepackage{amsmath}

\title[M.E.Bell et al]
{X-ray and radio variability in the low luminosity Active Galactic Nucleus NGC 7213}
\author[M.E.Bell et al]{M.E.Bell$^{1}$\thanks{E-mail:
meb1w07@soton.ac.uk (MEB)}, T.Tzioumis$^{2}$, P.Uttley$^{1}$, R.P.Fender$^{1}$, P.Ar\'{e}valo$^{3,4}$, E.Breedt$^{5}$, I.McHardy$^{1}$, 
\newauthor
D.E.Calvelo$^{1}$, O.Jamil$^{1}$ and E. K\"ording$^{6}$\\
$^{1}$Department of Physics and Astronomy, Southampton University, UK.\\
$^{2}$Australia Telescope National Facility, CSIRO, Marsfield, NSW, 2122, Australia.\\
$^{3}$Max-Planck-Institut f\"ur Astrophysik, Karl-Schwarzschild-Strasse 1, 85741 Garching, Germany. \\
$^{4}$Departamento de Ciencias Fisicas, Universidad Andres Bello, Av. Republica 252, Santiago, Chile.\\
$^{5}$Department of Physics, University of Warwick, Coventry, UK. \\
$^{6}$AIM/CEA Saclay and University Paris Diderot, l'Orme des Merisers, F-91219 Gif-sur-Yvette CEDEX, France.\\ }

\begin{document}

\date{Accepted 2010 September 13. Received 2010 July 01}

\pagerange{\pageref{firstpage}--\pageref{lastpage}} \pubyear{2010}

\maketitle

\label{firstpage}

\begin{abstract}

We present the results of a $\sim$ 3 year campaign to monitor the low luminosity active galactic nucleus (LLAGN) NGC 7213 in the radio (4.8 and 8.4 GHz) and X-ray bands (2-10 keV).  With a reported
X-ray Eddington ratio of $7 \times 10^{-4}$ L$_{Edd}$, NGC 7213 can be
considered to be comparable to a hard state black hole X-ray binary.  We show that a weak
correlation exists between the X-ray and radio light curves.  We use
the cross-correlation function to calculate a global time lag between
events in the X-ray and radio bands to be 24 $\pm$ 12 days lag (8.4
GHz radio lagging X-ray), and 40 $\pm$ 13 days lag (4.8 GHz radio
lagging X-ray).  The radio-radio light curves are extremely well
correlated with a lag of 20.5 $\pm$ 12.9 days (4.8 GHz lagging 8.4
GHz).  We explore the previously established scaling relationship
between core radio and X-ray luminosities and black hole mass $L_{R}
\propto M^{0.6-0.8}L_{X}^{0.6}$, known as the `fundamental plane of
black hole activity', and show that NGC 7213 lies very close to the
best-fit `global' correlation for the plane as one of the most luminous LLAGN.  
With a large number of quasi-simultaneous radio and X-ray observations, 
we explore for the first time the variations of a
single AGN with respect to the fundamental plane. Although the
average radio and X-ray luminosities for NGC 7213 are in good agreement with the plane, we show that
there is intrinsic scatter with respect to the plane for the individual data
points. 
 
\end{abstract}

\begin{keywords}
AGN Jets Radio X-Ray Lag
\end{keywords}

\section{Introduction}

The observable properties of active galactic nuclei (AGN) and black
hole X-ray binaries (BHXRBs) are consequences of
accretion on to a black hole at a variety of rates, in a variety of
`states', and within a variety of environments.  The major difference
between the aforementioned classes of object is the black hole mass.
BHXRBs typically have a black hole mass $\sim$10M$_{\odot}$ while for
AGN it is $10^{5}M_{\odot}\leq M \leq
10^{10}M_{\odot}$. Theoretically, the central accretion processes
should be relatively straightforward to scale with mass, and this is
supported by several observed correlations. These include a relation
between the X-ray and radio luminosities and the black hole mass
(Merloni, Heinz \& di Matteo 2003; Falcke, K\"ording \& Markoff 2004),
and between X-ray variability timescales, mass accretion rate and mass
(McHardy et al. 2006). More quantitative similarities between
accretion `states' and radio jet production have also been
demonstrated (K\"ording, Jester \& Fender 2006; for the current
picture of accretion states in BHXRBs and their relation to radio jets
see Fender, Belloni \& Gallo 2004).

Studying the delays between different emission regions gives us a
further handle on the scalability of black hole accretion, as signals
propagate from, for example, the accretion flow to the jet.
Variability studies have so far shown that a correlation exists
between the X-ray and optical emitting regions of both BHXRBs and AGN, typically 
reporting small lags, which are consistent with at least some of the optical variations being due to X-ray heating of the disc \citep{Dave,Elme}. 
A recent study by \cite{PG} has shown that a correlated time lag of
$\sim$ 100 ms exists between the X-ray and IR regions (IR lagging X-rays) for the BHXRB
GX339-4, indicating a close coupling between the hot accretion flow
and inner regions of the jet.  In the case of the BHXRB GRS 1915+105 a
variable X-ray to radio lag of $\sim 30$ mins (radio lagging X-ray) has been measured 
\citep{Pooley_Fender,Fender_1999}.  Discrete ejection events have been
resolved in both the AGN 3C120 \citep{Marscher} and GRS 1915+105
\citep{Pooley_Fender,Fender_1999}.

The linear scaling with mass of the characteristic timescale around a
black hole means that there are advantages to studying each class of
object. In BHXRBs we can track complete outburst cycles, from the
onset of disc instabilities through major ejection events, radio-quiet
disc-dominated states, and a return to quiescence, on
humanly-observable timescales (typically years). For a typical AGN the
equivalent cycle may take many millions of years. However, for an AGN
we are able to resolve individual variations of the source on
time-scales that are comparable to or shorter than the shortest physical time-scales in the system (e.g. the dynamical time-scale), something
which is currently impossible for BHXRBs. In `black hole time' we are
able to observe the evolution of sources in fast-forward for BHXRBs
and in detailed slow-motion for AGN.

In this paper we present the results of a long term ($\sim 3$ years)
regular monitoring campaign in the X-ray and radio bands of the low
luminosity active galactic nucleus (LLAGN) NGC 7213. 
Previous X-ray studies show that NGC 7213 is accreting at a low rate $\sim 7 \times 10^{-4}$
L$_{Edd}$ \citep{Starling}. 
The hard state in BHXRBs is typically observed at bolometric luminosities below $\sim 1\%$ Eddington, and
seems to be ubiquitously associated with a quasi-steady jet.  Above
$\sim 1\%$, sources can switch to a softer X-ray state, the jets are
suppressed \citep{Tom_Ed,Dunn}; furthermore transition to this softer
state is usually associated with major transient ejection events.
As NGC 7213 is considerably below L$_{Edd}\sim$ 1\% we therefore consider 
it a good candidate for comparison with other BHXRBs in the low/hard state.
If we consider AGN to be `scaled up' versions of BHXRBs by exploring the
time lag between the X-ray and radio emitting regions we can compare, contrast and hopefully relate the accretion and jet production
scenarios for AGN and BHXRBs.

A correlation has been established by \cite{Corbel_2003} and Gallo et
al. (2003, 2006) relating the radio luminosity ($L_{R}$) and X-ray
luminosity ($L_{X}$) for BHXRBs in the low/hard and quiescent states,
where $L_{R} \propto L^{0.6}_{X}$.  
\cite{Merloni} - hereafter MHdM03 and \cite{Falcke_2004} extended the
BHXRB relationship using two samples of AGN to form the `fundamental
plane of black hole activity'. By accounting for the black hole mass
(M) the relationship $L_{R} \propto L^{0.6}_{X}$ has been extended to
cover many orders of magnitude in black hole mass and luminosity.  Further refinements were made to the fundamental plane by
\cite{KFC} - hereafter KFC06, using an augmented and updated sample to
examine the fitting parameters.

Throughout this paper we define the `intrinsic' behaviour of AGN and
BHXRBs as multiple measurements (in the radio and X-ray) of the {\em
  same} source. We define the `global' behaviour as single (or average)
measurements of {\em multiple} sources, both with respect to the
fundamental plane.  

For the BHXRBs in the low/hard state the
relationship described above has not only been established globally
but in some cases intrinsically, i.e GX 339-4, V404 Cyg and a small
number of other systems have been shown to move up and down the
correlation seen in
the fundamental plane \citep{Fender}. However, in recent years an
increasing number of outliers have been found below the correlation,
i.e. less radio-loud then expected  (\cite{Xue_XRB}; \cite{Gallo_2007}; \cite{Dan}) 
as well as some sources which move in the plane with a
different slope (e.g \cite{Jonker}).
To date the correlation found from the fundamental plane has only been measured
globally for AGN, not intrinsically. Note, with respect to the global
measurements of the AGN population, the specific measurements of the
radio and X-ray flux used in the correlation are sometimes taken at
different times and thus could be a source of error in the correlation
\citep{KFC}.

As well as establishing the time lags, another goal of this
work, is to establish, through quasi-simultaneous observations the
intrinsic relationship between $L_{R}$, $L_{X}$ and M observed in an
LLAGN and its relevance to the fundamental plane of black hole
activity.  We use the MHdM03 and KFC06 samples for comparison both
with an updated BHXRB sample taken from Fender, Gallo \& Russell
(2010) - hereafter FGR10.  We also explore the possible scatter in AGN data points
away from the fundamental plane and place limits on this deviation.

\section{NGC 7213 Background}

NGC 7213 is a face-on Sa galaxy hosting a Seyfert 1.5 nucleus located
at a distance of 25 Mpc assuming H$_{0}$=71kms$^{-1}$Mpc$^{-1}$. 
Narrow low ionisation emission lines are observable in its nuclear spectrum, 
also making it a member of the LINER class \citep{Liner_ref}.

The radio properties of NGC 7213 are intermediate between those of radio-loud
and radio-quiet AGN.  Previous radio studies at 8.4 GHz have not
resolved any jet emission from the nucleus on scales 3 mas to 1 arcsec
(\cite{Blanford_ACTA}, \cite{Thean_High_res},\cite{Blank_LBA}).The Long
Baseline Array (LBA) is an Australian six station VLBI instrument \citep{LBA}; LBA observations at 1.4 and 0.843 GHz have shown some
evidence for larger scale emission that could be due to jet-fed radio
lobes \citep{Blank_LBA}. However, as the nucleus remains unresolved it could be
that the jet is oriented to some degree in the direction of the observer. As
suggested by \cite{Blank_LBA}, the lower frequency emission could
possibly be associated with a `kink' in the jet at a larger distance.
This could be consistent with a general model proposed by
\cite{Falcke_1996} where radio-intermediate objects are in fact radio
-quiet objects whose jets are to some extent aligned in the observers
direction and relativistically boosted.

Previous X-ray studies have so far failed to show significant evidence
for a soft X-ray excess; Compton reflection; or a broad Fe K$\alpha$
line in the NGC 7213 spectrum e.g. see \cite{Bianchi_2003},
\cite{Bianchi_2008}, \cite{Starling}.  A narrow Fe K$\alpha$ line is
observed and \cite{Bianchi_2008} show evidence that it is produced in
the broad line region (BLR). \cite{Starling} also report that the
expected UV bump is either absent or extremely weak.  The weakness or absence of these signatures suggests that the inner accretion disc is missing, perhaps replaced by an advection dominated accretion
flow (ADAF) (as suggested by
\cite{Starling}) or similar hot flow, consistent with the low/hard state interpretation of NGC~7213.

\cite{Blank_LBA} estimate the black hole mass of NGC 7213 using the
nuclear velocity dispersion calculation of \cite{Nelson} and the
velocity-dispersion versus black-hole-mass relationship of \cite{Ferrarese}
to obtain $9.6^{+6.1}_{-4.1}\times10^{7} M_{\odot}$.

\section{Observations}
\subsection{ATCA data analysis}
Bi-weekly monitoring was obtained with the Australia Telescope Compact Array (ATCA).
The interferometer setup was such that 128 channels of 1 MHz bandwidth were used to form two continuum channels centred at 4.8 and 8.4 GHz respectively. 
The radio observations have been reduced using the MIRIAD radio reduction package \citep{Sault}. 
Flux and bandpass calibration was achieved using (in most cases) PKS J1939-6342 (B1934-638) and for the phase calibration PKS J2218-5038 (B2215-508).

A variety of fitting techniques were then tested to extract the flux density of the source from the observations. 
These included testing point and Gaussian fitting to the source in the image and uv plane. 
As ATCA is an East-West array and observations were short in duration ($\sim$2 hours) an elongated synthesised beam was typically produced. 
Image-plane fitting often failed to converge and/or produced non-physical results.
Therefore fitting was discarded in the image plane for the more reliable uv plane method. 
The final radio light curves and spectral indexes (between the radio bands only) are shown in figure \ref{lightcurve}.

To complement the snapshot observations a full 12 hour integration was performed on 2008 June 29; the purpose of which was to explore the polarisation properties of the jet, and to detect any faint extended structure. Additional steps were taken to ensure adequate polarisation calibration was performed.
This observation was also used to make a deep image of the source; beyond the 0.5$\arcsec$ beam no weak extended radio emission was observed above a noise level $\sigma$=78$\mu$J at 4.8 GHz.

\subsection{RXTE data analysis}
\begin{figure}
\centering
\includegraphics[scale=0.58]{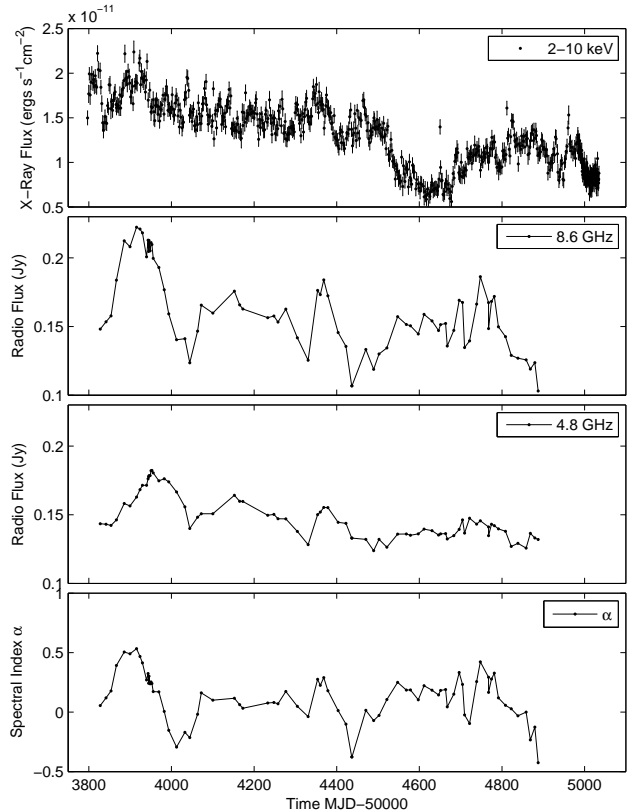}
\caption{Top Panel: X-ray Flux verus Time (MJD-50000). Center Top: Radio flux $S_{v8.4 GHz}$ versus time. Center Bottom: Radio flux $S_{v4.8 GHz}$ versus time. The errors are too small to show on both radio flux figures but are typically $\sim$0.4 mJy/Beam. Bottom: Spectral index (8.4-4.8 GHz) $\alpha$ versus time where $S_{v}\propto\nu^{+\alpha}$.}
\label{lightcurve}
\end{figure}

Daily monitoring was obtained using 1~ksec snapshots obtained with the Rossi X-ray Timing Explorer (RXTE) Proportional Counter Array (PCA), allowing a long-term light curve to be obtained in the 2-10 keV band. 
The data were reduced using FTOOLS v6.8, using standard extraction methods and acceptance criteria. The background was calculated from the most recent background models which corrects for the recent problems with the $RXTE$ SAA (South Atlantic Anomaly) history file. The final 2-10 keV fluxes were calculated by fitting a power law to the observed spectra. This allows us to take into account changes in the instrumental gain over the duration of the monitoring. The final un-binned X-ray light curve is shown in figure \ref{lightcurve} and is used in all subsequent analysis.

\section{Results}
\subsection{X-ray/radio light curves and cross-correlation} 
Figure \ref{lightcurve} shows X-ray flux (top panel), radio flux (middle panels) and radio spectral index $\alpha$ 
where $S_{v} \phantom{i}\propto\phantom{i}\nu^{+\alpha}$ (bottom panel) versus time . 
The X-ray light curve shows a general decrease in flux until MJD 54600. Two distinct X-ray flares are observed at MJD 53920 and MJD 54390, both appear to be correlated with events in the radio bands. 

We use the Discrete Correlation Function (DCF) method of \cite{CCF_Paper} to calculate the cross-correlation coefficients between the entire X-ray and radio bands to find the lag. 
To calculate the centroid lag $\tau_{cent}$, we use a weighted mean of the positive lags above an 85\% threshold of the peak DCF value. We use $\tau_{cent}$, rather than the peak value $\tau_{peak}$, because the centroid has been shown to better represent the physical lag \citep{Koratkar}. It has been shown that it is difficult to interpret $\tau_{peak}$ as a physical quantity, and if it is used to calculate the lag, it usually offers an underestimate when compared with $\tau_{cent}$ \citep{Peterson}. The centroid width is calculated as  

\begin{table*}
\centering
\caption{Summary of the lag times which have been calculated using the Discrete Correlation Function for the X-ray-8.4 GHz and X-ray-4.8 GHz light curves. We used the z-transformed discrete correlation function to calculate the lag for the 8.4 GHz-4.8 GHz and flare 1 \textit{only} light curves.}
\begin{tabular}{|c|c|c|c|c|}
\hline
Time (MJD) & X-ray-8.4 GHz Lag (days) & X-ray-4.8 GHz Lag (days) & 8.4 GHz-4.8 GHz Lag (days) & \\
\hline
\hline
All & $24 \pm 12$ (DCF$_{max}$=0.56) &$ 40 \pm 13$ (DCF$_{max}$=0.7) & $20.5 \pm 12.9$ (ZDCF$_{max}$=0.81) & \\
Flare 1 (53800-54000) & $14 \pm 11$ (DCF$_{max}$=0.78)  &$ 35 \pm 16$ (DCF$_{max}$=0.73) & - & \\
\hline
\label{lag_table}
\end{tabular} 
\end{table*}

\begin{equation}
\tau_{cent} = \frac{1}{N}\sum_{i}\frac{c_{i}\tau_{i}}{c_{i}}
\end{equation}
Where $c_{i}$ and $\tau_{i}$ are the DCF coefficients and lags, $\tau_{cent}$ is the centroid lag. As the DCF coefficients are only defined for a given lag, we interpolate to find the peak correlation coefficient DCF$_{max}$ at the centroid lag .

\begin{figure}
\centering
\includegraphics[scale=0.78]{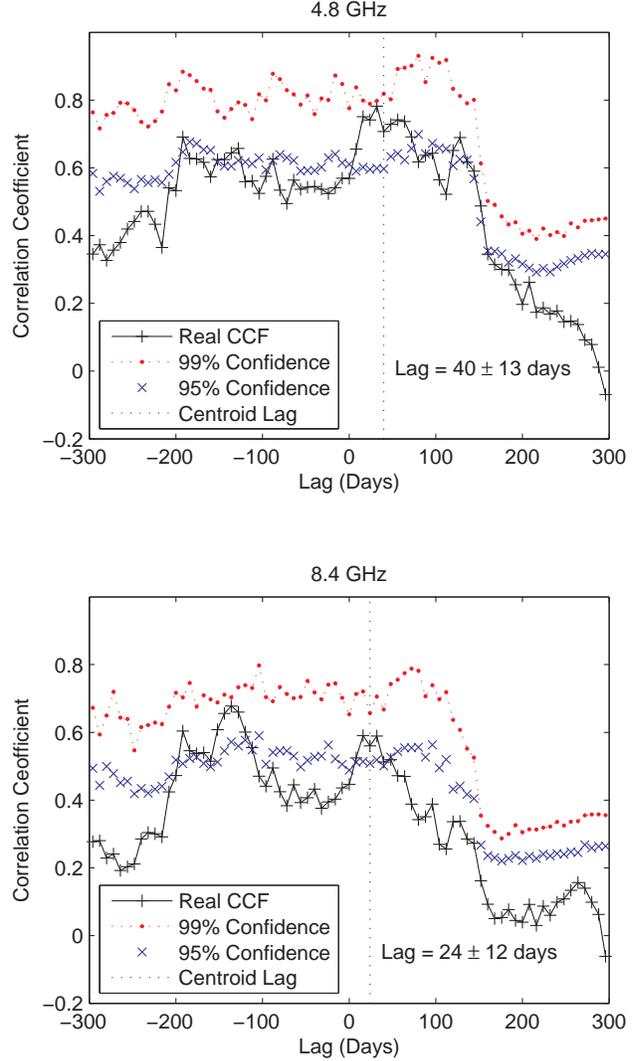}
\caption{Top Panel: Discrete cross-correlation function (DCF) plot showing time lag against correlation coefficient for X-ray to radio (4.8 GHz lagging X-ray) (black crosses). A lag of 40 $\pm$ 13 days at DCF$_{max} = 0.7$ is calculated using the weighted mean of the lags above an 85\% threshold of the peak DCF value. 
Bottom Panel: Discrete cross-correlation function (DCF) plot showing time lag against correlation coefficient for X-ray to radio (8.4 GHz lagging X-ray) (black crosses). A lag of 24 $\pm$ 12 days at DCF$_{max} = 0.56$ is calculated using the weighted mean of the lags above an 85\% threshold of the peak DCF value.  On both plots the 99\% and 95\% local significance confidence levels are plotted. The errors are calculated via the method described in section 4.1} 
\label{CCF_both}
\end{figure}

Figure \ref{CCF_both} show the DCFs for the X-ray-8.4 GHz and X-ray-4.8 GHz bands respectively.
We find a lag of $\tau_{cent}=$24$\pm$12 days at DCF$_{max}$=0.56 for X-ray-8.4 GHz and $\tau_{cent}=$40$\pm$13 days at DCF$_{max}$=0.7 for X-ray-4.8 GHz.
The time lags taken from figure \ref{CCF_both} are summarised in table \ref{lag_table}. 
The errors in the lag are calculated using the flux randomisation/random subset selection (FR/RSS) method of \cite{Peterson}. For $10^{4}$ samples, we randomly reduce the number of data points in our original X-ray and radio light curves by 37\% and then re-calculate the DCF. We take the centroid lag from each of the 10$^{4}$ simulations and use the rms spread in the distribution of centroid lags to obtain the quoted errors.  Note that since the radio light curves are strongly correlated with one another, the statistical errors on the lags are not independent between the two bands, so that the difference in lag between the two radio bands with respect to X-rays is likely to be real (this is also highlighted by the radio-radio correlation, see below).

\begin{figure}
\centering
\includegraphics[scale=0.6]{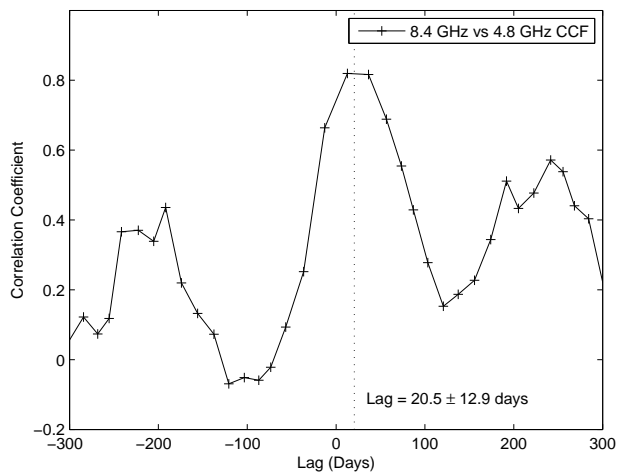}
\caption{The z-transformed discrete cross-correlation (ZDCF) function between the 8.4 GHz and 4.8 GHz light curves(4.8 GHz lagging 8.4 GHz). A centroid peak of 20.5 $\pm$ 12.9 days is found at ZDCF$_{max}$=0.81.}
\label{CCF_radio}
\end{figure}

To ascertain a confidence level in the DCF calculation we used the method of \cite{Timmer} to generate Monte-Carlo simulations of un-correlated red-noise light curves. We assumed a broken power-law-shape power spectrum with low-frequency slope -1, high-frequency slope of -2.3 and break frequency $\nu=0.012$~d$^{-1}$, based on the best-fitting model to fit the X-ray power spectrum (Summons, private communication). We used this model to generate $10^{4}$ random X-ray light curves and cross-correlated them with the real radio light curves in both bands. We then used these correlations to estimate the distribution of cross-correlation values at \textit{each} lag to obtain a local confidence level, corresponding to the CCF value which exceeds $P$~per-cent of the simulated CCF values at that lag.
From the simulations we plot on figure 2 the $95\%$ and $99\%$ local confidence levels: we therefore assign a local confidence of $>95\%$ for both of the peaks with the 4.8 GHz peak reaching close to 99$\%$. 

It is important to note however, that without a priori expectation of what the lag should be, it is more statistically rigorous to estimate the significance of the correlation using the `global significance', which is the fraction of simulated random light curves which show CCFs at better than the observed confidence level for {\it any} given lag. For example, to estimate the global significance of a peak which appears at the 99$\%$ local confidence level, we search the $10^{4}$ simulated CCFs for any peak values (at any lag) which exceed the 99$\%$ local confidence level for that lag.  The fraction of CCFs which do not show such a peak defines the global significance level for the correlation observed at the given local confidence level.  In this way, we account for the fact that we are effectively searching over a broad range in lags, and so are sampling from a larger population of potential `false positives' than we would expect from searching for correlations at a single lag.

We find for the entire range of computed lags (-300 to +300 days) the global confidence levels at 4.8 GHz are 82.0\% (at 99\% local significance) and 56.2\% (at 95\% local significance); at 8.4 GHz, 77.1\% (at 99\% local significance) and 46.0\% (at 95\% local significance).  If we restrict the range of lags to positive only ($>$ 0 days) we find at 4.8 GHz, 91.0\% (at 99\% local significance) and 74.7\% (at 95\% local significance); at 8.4 GHz, 86.6\% (at 99\% local significance) and 63.3\% (at 95\% local significance). 

We note that in the case where we do not restrict the lag to lie within any given range, the global significance is low.  For example, it is not surprising that the 8.4~GHz versus X-ray CCF shows two peaks at greater than 95$\%$ local confidence, since the corresponding global confidence at this level is only 56$\%$, so we expect a spurious peak in nearly half of the CCFs sampled from random light curves.  However the peak corresponding to a negative lag of $\sim140$~days (i.e. radio leading X-ray) cannot easily be explained with any physical model. When we presume that there must be a positive lag (radio lagging X-ray) - as we would indeed predict based on our current understanding of the accretion disk/jet - this global significance level increases.  Put another way, it is statistically unlikely that a spurious correlation should appear in the range of lags which matches our physical expectations, as seems to be the case here. Therefore, we assume for the remainder of the paper that the correlation is real.  Note that if the correlation is real, the lags can be well-determined: the significance of a lag and the significance of a correlation itself are unrelated quantities, since a lag can be very well determined from just a single well-sampled `event' (flare or dip) in red noise data, if the data are indeed correlated.  We will explore possible complexities in the disk/jet connection  with respect to our calculated lags in section 5.1  

\begin{figure}
\centering
\includegraphics[scale=0.75]{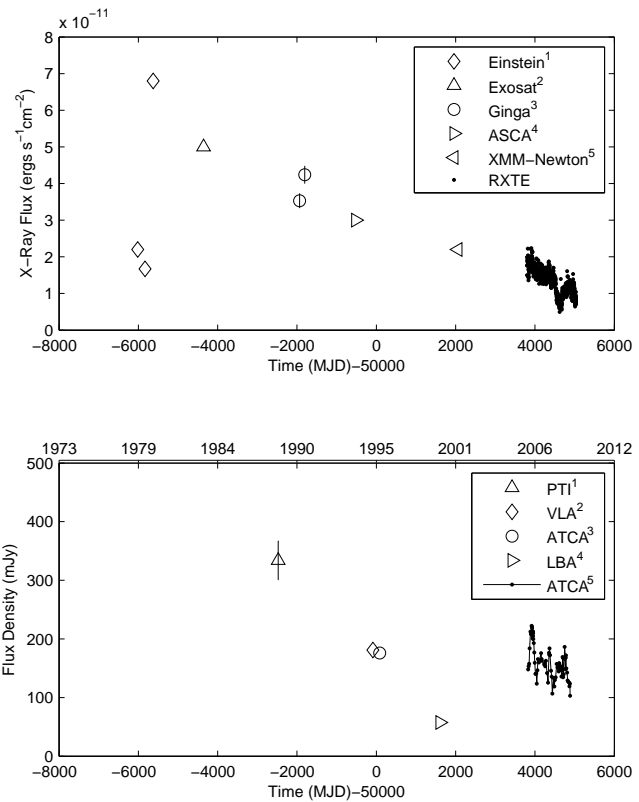}
\caption{Top Panel: Light curve of archival X-ray data points in the 2-10 keV range. Flux values cited in $\diamond$1=\citep{Einstein_1}, $\triangle$2 = \citep{Turner}, $\circ$3 =\citep{Nandra}, $\triangleright$4=ASCA using the tarturus database \citep{Tartarus}, $\triangleleft$5 = \citep{Bianchi_2008}. Bottom Panel: Light curve of archival radio data points at 8.4 GHz. Flux values cited in $\triangle$1 = \citep{Slee}, $\diamond$2=\citep{Thean_High_res}, $\circ$3 = \citep{Blanford_ACTA} and $\triangleright$4 = \citep{Blank_LBA}}
\label{archiveplot}
\end{figure}

\subsection{Radio/radio cross-correlation}
To calculate the cross-correlation between the two radio bands we used the z-transformed discrete correlation function (ZDCF) of \cite{Alexander}, see figure \ref{CCF_radio}. Note that the two radio light curves are sampled at the same time: as noted in \cite{CCF_Paper}; to keep the normalisation correct for the DCF we need to omit the zero lag pairs as the errors between the two bands are correlated. Therefore for convenience (and as the two methods are comparable) we switch to the ZDCF which omits zero lagged pairs. We calculate the errors using the same method as described in section 4.1.
We thus find $\tau_{cent}=20.5 \pm 12.9$ days which is consistent with the difference in lags between X-ray-4.8 GHz and X-ray-8.4 GHz. 

\subsection{X-ray/radio cross-correlation when flaring}
We split the light curve up into data points surrounding the first flare at MJD 53920, which is summarised in table \ref{lag_table}. We then calculated the ZDCF to ascertain an accurate measurement of the lag for this flare $only$. We find $\tau_{cent}=14 \pm 11$ days for X-ray-8.4 GHz, and $\tau_{cent}=35 \pm 16$ days for X-ray-4.8 GHz. Both lag times are below the global lag for the entire light curve. We do not perform this analysis for the second flare at MJD 54380 as it was only sparsely sampled in the radio. 

\subsection{Long term variability}
The long term X-ray variability, covering a time-range of about 30 years, is plotted in the upper panel of figure \ref{archiveplot}. Archival data were searched for in the literature from the instruments $Einstein$ \citep{Einstein_1}, EXOSAT \citep{Turner}, $Ginga$ \citep{Nandra}, ASCA using the Tartarus database \citep{Tartarus} and $XMM-Newton$ \citep{Bianchi_2008}. Around MJD 44400 ($\sim$ 1980) a sharp flare which is brighter than the flares visible in our dataset was observed. Since then the flux has been gradually decreasing.

\begin{table*}
\centering
\caption{Summary of archival observations found in the literature. The data points between the horizontal lines indicate where the two fluxes were used to calculate the X-ray-radio correlation. Two X-ray and two radio points were taken at similar times thus they were both used to calculate the correlation, (R) and (X) denotes the time (MJD) of the observation for radio and X-ray respectively. $\delta$t gives the time difference between observations (Radio-X-ray). The instrument used to derive the flux is given in brackets (see section 4.4 for references) and errors are given only when reported in the literature.}
\begin{tabular}{|c|c|c|c|c|}
\hline
Time (MJD) & Radio Flux $S_{v}$ 8.4 GHz (mJy) & X-ray Flux 2-10 keV (ergs$^{-1}$) & $\delta$t$_{meas}$ (Radio - X-ray) (days) & \\
\hline
\hline
43985 &-& 2.2$\times10^{-11}$ (Einstein) & - & \\
44168 &-& 1.7$\times10^{-11}$ (Einstein)& - & \\
44374 &-& 3.8$\times10^{-11}$ (Einstein)& - & \\
45644 &-& 5.0$\times10^{-11}$ (Exosat)& - & \\
\hline
47527 (R) - 48065 (X) & 334$\pm$33 (PTI)   &  3.5$\times10^{-11} \pm 0.2\times10^{-11}$ (Ginga)& -538 & \\
47527 (R) - 48193 (X) & 334$\pm$33 (PTI)   &  4.2$\times10^{-11} \pm 0.25\times10^{-11}$ (Ginga)& -666 & \\
\hline
49913 (R) - 49479 (X) & 181 $\pm$1 (VLA) & 3.0$\times10^{-11}$ (ASCA) & 434 & \\
50083 (R) - 49479 (X) & 176 $\pm$1 (ATCA)& 3.0$\times10^{-11}$ (ASCA)& 604 & \\
\hline
51604 (R) - 52057 (X) & 57.6 $\pm$1.3 (LBA) & 2.2$\times10^{-11}$ (XMM-Newton)& -453 & \\
\hline
\label{Achival_Table}
\end{tabular} 
\end{table*}

The long term radio variability at 8.4 GHz is plotted in the lower panel of figure \ref{archiveplot}. The flux values were taken from a variety of instruments and publications (which are referenced on the plot).
In order of increasing time the first point on the plot was taken from \cite{Slee} using the Parkes-Tidbindilla interferometer (PTI). 
Then near simultaneous data points were taken with the VLA \citep{Thean_High_res} and then ATCA \citep{Blanford_ACTA}. 
The flux from the long baseline array (LBA) observation by \cite{Blank_LBA} was the last point plotted prior to the ATCA monitoring which is presented in this paper. 

When comparing the flux expected from the start of our ATCA monitoring with the LBA data point, the LBA point seems too low.
The core is un-resolved with all of the radio telescopes described above. 
The PTI observations constitutes the first data point in figure \ref{archiveplot} and the LBA the last (before ATCA monitoring).
Note, the Parkes-Tidbindilla baseline is part of the LBA network. 
\cite{Blank_LBA} comment that in the LBA observations, no decrease in flux was observed with respect to baseline length i.e it was not resolved; they conclude from this that the decrease in flux with respect to the other observations is accurate.

However, both the PTI and the LBA are not sensitive to the same spatial resolutions as the VLA and ATCA (the largest ATCA baseline is 6km and the shortest LBA baseline used in the archival observation is 113km). 
It is possible that some flux (on ATCA equivalent baselines) is not accounted for: which could plausibly give a decrease in flux i.e indicating brighter, larger scale structure.
The logical consequence of this is that the PTI data point is also missing some flux as it samples on a singular baseline of 275km. 
We will discuss these measurements within the context of the fundamental plane and attempt to draw some conclusions in section 5.2

\subsection{Linear polarisation}
A full 12 hour synthesis ATCA observation was performed on 2008 June 29 at 8.4 and 4.8 GHz. The purpose of this observation was to perform a reliable polarisation calibration to ascertain an accurate measurement of the percentage linear polarised (LP) flux from the source. 
\cite{Percent_Flux_LLAGN_Bower} showed that from a sample of 11 LLAGN the mean LP was $\sim 0.2 \%$. \cite{ATCA_Blazar} from a survey of 22 blazars report a mean LP $\sim 3\%$ with a standard deviation $\sigma=1.5$.
At the time of observation the polarisation of NGC 7213 was $<0.1\%$ at 4.8 and 8.4 GHz respectively, which is more consistent with the reported value of typical LLAGN and not blazars. The percentage polarisation was calculated at other epochs, however without a full 12 hour synthesis parallactic angle calibration the values were often suspect and will not be presented here. 

\begin{figure*}
\centering
\hspace{-20mm}
\includegraphics[scale=0.8]{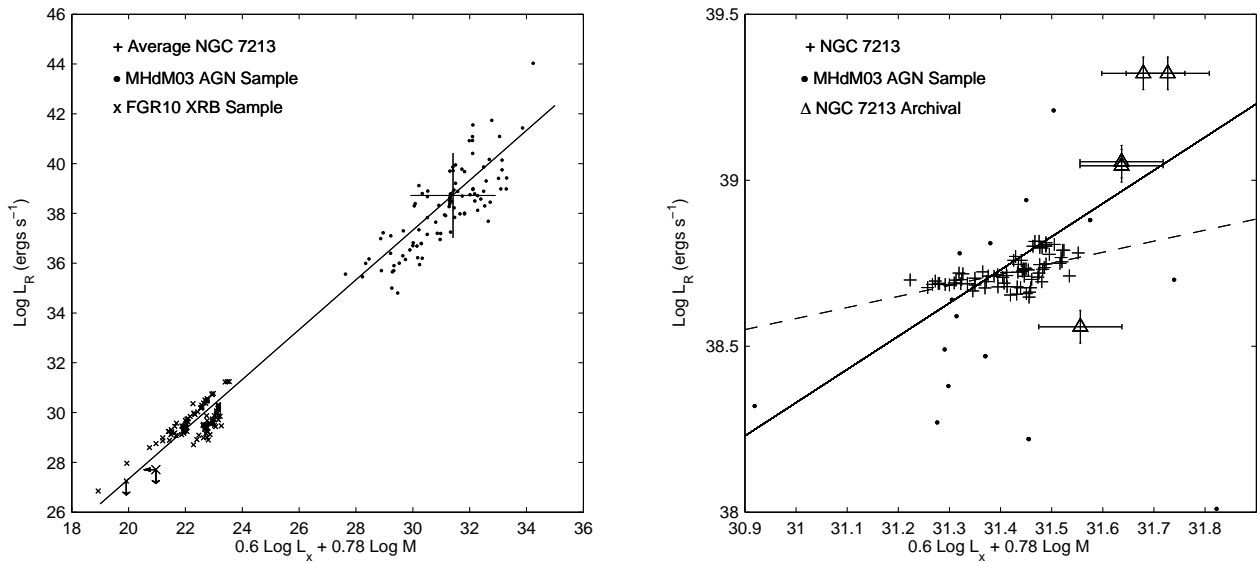}
\caption{Left Panel: The fundamental plane of black hole activity. The large cross indicates the average location of NGC 7213. Black dots are the MHdM03 sample with the black line indicating the best fit. Crosses indicate the updated BHXRB sample of FGR10; data points have been converted into the 2-10 keV energy range. Right Panel: 
A close up of the NGC 7213 data points with least squares best fit indicated with a dashed line. The MHdM03 fit is shown with a solid line. The archival data points are also shown with but are not used in the NGC 7213 data best fit. NGC 7213 measurements are taken at 4.8 GHz and the archival are at 8.4 GHz.}
\label{full_plane}
\end{figure*}

\section{Discussion}
\subsection{X-ray/radio jet connection}

We have shown that a weak but statistically significant delay exists between the X-ray and radio emitting regions, with the radio lagging behind the X-ray. A number of models have been proposed to explain and interpret the X-ray to radio lag in both BHXRBs and AGN. The most notable of these are the `internal shock model' \citep{Bland_79,Rees} and a `plasmon model' \citep{vanderlaan}. We will briefly explore these models and comment on the relevance - if any - to our data.

We first consider a plasmon model.
After the accretion mechanism(s) have pushed matter into the jet, an adiabatically expanding initially self-absorbed synchrotron emitting plasmon travels at relativistic speeds down the path of the jet. At a given time and frequency the matter becomes optically thin and is `detectable' at that given frequency (in this case either 8.4 and 4.8 GHz). If the jet is fed at a constant mass rate, density and velocity the delay time for material to become optically thin will be constant.
If these parameters are not constant the delay between X-ray and radio will be variable (e.g. see \cite{vanderlaan}) assuming that the disk and the jet are indeed coupled \citep{SagA_Falcke}. Note, the higher frequency radio emission (8.4 GHz) will become optically thin first, thus a delay from 8.4 to 4.8 GHz is always expected.

Another model for explaining the emission seen in jets is the `internal shock model' \citep{Bland_79,Rees}.
The synchrotron lifetime of an emitting region is too short to adequately explain the scales of jets observered in AGN.
These emitting regions - commonly referred to as `knots' - are often displaced from the central core emission. 
Localised shocks within these regions are needed to explain the time-scales of variation observed \citep{Rees, Felton}.
Jet shock scenarios have also been used to model the common flat spectrum jet observered in BHXRBs and AGN (e.g. see \cite{Spada,Omar}).

Figure \ref{lightcurve} shows the evolution of spectral index; indicating that during a flare event $\alpha$ increases ($\alpha>0$), flattens and steepens ($\alpha<0$) shortly afterwards.
The initial re-energisation given by a shock would push/compress the plasma into the optically thick regime; subsequently this process moves the material into the optically thin regime.
The variable time lag seen in the DCF could be interpreted as the time taken for newly injected matter to `catch up' with older matter expanding adiabatically in the jet to shock and produce a flare. The lag would be dependent on the Lorentz factor of the newly injected material which in turn is related to the accretion rate \citep{SagA_Falcke}. 

Although both of these models can adequately explain the variations observed in our light curves, we cannot completely disentangle them.
It is possible that both of these models are in some part responsible for the behaviour.
As the jet remains unresolved we cannot identify an area of localised shocks or indeed discrete resolved events that we could use to support a particular model.

To explore - in a very simplistic sense - how the time lags seen in
BHXRBs scale up to AGN we offer two comparisons; that of a simple mass
scaling and also mass-Eddington ratio scaling.  For example, discrete
resolved ejection events have been seen in GRS 1915+105 with a
variable time lag between X-ray and radio (GHz), with the clearest
examples of events having lags of $T_{GRS1915}=$ 20-30
mins \citep{Pooley_Fender,Mirabel}.  Taking the characteristic
timescale of X-ray-radio lags measured in GRS 1915+105 and scaling up
to NGC 7213 with mass only (assuming the mass of GRS 1915+105 as
$\sim$ 10M$_{\odot}$ and the mass of NGC 7213 as 9.6 $\times
10^{7}$M$_{\odot}$ and using $T_{lag} = T_{GRS1915}\times
(M_{NGC7213}/M_{GRS1915})$) we infer $T_{lag}=$ 2 $\times 10^{5}$
days, much longer than we observe. GRS 1915+105 is however accreting at
a much higher rate,
therefore including a scaling by the ratio of the Eddington
luminosities (using $T_{lag2} = T_{lag}\times
(L_{Edd-NGC7213}/L_{Edd-GRS1915})$ and taking $L_{Edd-GRS1915} \sim
1$) we find $T_{lag2}=$ 140 days, reasonably comparable to our measured
lag. 

The difficulty in comparing the
actual X-ray-radio time lag measured for NGC 7213, and that of the GRS
1915+105 scaled time lag is that the radio emitting regions being
probed are significantly different: in a broader view the spectral
energy distributions are different. In fact until the structure of jets
and how it scales with accretion rate and mass at a given frequency are
better understood, such attempts at quantitative comparison are of
marginal value, although the qualitative comparison is valuable.

\begin{figure}
\centering
\includegraphics[scale=0.75]{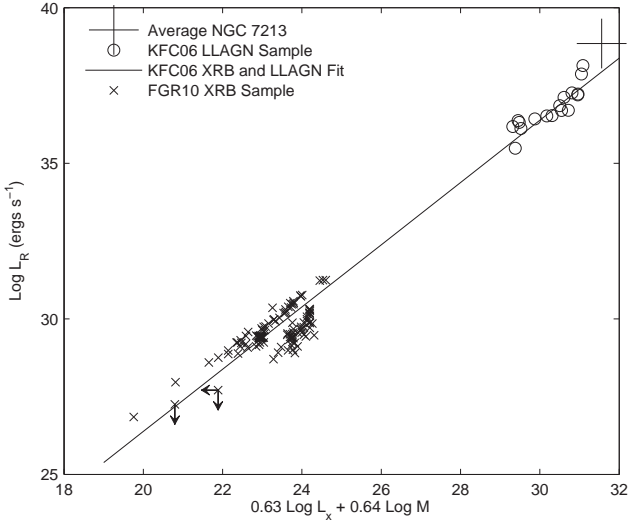}
\caption{The fundamental plane using the KFC06 LLAGN sample only (circles) and updated FGR10 BHXRB sample (crosses). The large cross indicates the average location of NGC 7213. The solid line indicates the KFC06 best fit.}
\label{KFC}
\end{figure}

\subsection{The fundamental plane of black hole activity}
The fundamental plane of black hole activity \citep{Merloni,Falcke_2004} shows that the correlation $L_{R} \propto M^{0.8}L_{X}^{0.6}$ holds over many orders of magnitude of X-ray and radio luminosities, and black hole masses. 
Figure \ref{full_plane} shows; at high luminosities a sample of AGN taken from MHdM03, and at the low luminosities an updated BHXRB sample taken from FGR10.
The updated BHXRB X-ray data were observed in the 0.5-10 keV energy range, therefore careful steps were taken (using WebPIMMS) to correctly convert into the 2-10 keV range to compare with the MHdM03 and NGC 7213 data points.
We plot on figure \ref{full_plane} the best fit parameters $L_{R} = (0.6^{+0.11}_{-0.11})$ log $L_{X}+(0.78^{+0.11}_{-0.09})$ log $M+7.33^{+4.05}_{-4.07}$ as defined by MHdM03. Note, we do not re-fit the best fit line for the updated BHXRB sample. 

We paired the X-ray and radio data points to calculate the correlation by finding the closest X-ray point (in time), to the radio, because the X-ray sampling was more frequent.
The average for NGC 7213 sits well on the predicted correlation. 
The right hand panel of figure \ref{full_plane} shows a close up of the NGC 7213 data.
A least squared best fit for the NGC 7213 data points only is shown and parametrised by $L_{R} = (0.2^{+0.056}_{-0.056})$ log $L_{X}+28.3^{+1.77}_{-1.77}$ (the constant includes the mass term in the context of MHdM03). 
 
We include in figure \ref{full_plane} the correlation found from the archival X-ray and radio data; however, as they were not simultaneous we do not include them in the best fit. 
We paired the X-ray and radio data points to calculate the correlation according to the closest in time. If either two X-ray, or two radio points were very close in time, both are included in the correlation. 
Also note that these data were calculated between 8.4 GHz and 2-10 keV and not 4.8 GHz. Assuming that the same correlation (i.e $\sim$ flat spectrum) holds between the 8.4 and 4.8 GHz observations the plot seems to show that two of the archival points do not follow the trend predicted by the fundamental plane. 

To assess whether the deviation/scatter in the archival points (and indeed the other MHdM03 AGN) away from the fundamental plane can be explained due to a delay in sampling the X-ray and radio fluxes; 
we take the fraction RMS scatter in the RXTE/ATCA X-ray and radio data and plot this errorbar with the archival points. 
We find the fractional X-ray RMS to be $fRMS_{x}$=26\% and the fractional Radio RMS $fRMS_{r}$=11\%.
The full light curve presented in this paper spans $\sim$ 1000 days and the longest change in time between the archival observations is $\sim$ 500 days (see table \ref{Achival_Table}). Although this method only offers an estimate of the error between sampled data points, it does allow us to assess outliers.  

Within this framework it appears that the first (PTI correlation) and last (LBA correlation) data points that were used in the correlation are outliers: the ATCA and VLA correlation points sit close to the best fit. Even when taking into account the full RMS from the 1000 day light curve the error bars do not bring the data points close to the MHdM03 best fit.
We have also speculated earlier that there could be a certain amount of missing flux associated with these long baseline observations. 
Moving both the PTI and LBA points up by some set amount will still leave the PTI point away from the correlation. Although applying the intrinsic variation of the source to the archival points cannot bring all points onto or close to the correlation; the scatter in these points is within that permitted by the other MHdM03 points shown on the plot. 

By using the fractional RMS of X-ray and radio variability from our study we have placed a constraint in the deviation of the archival data points away from the fundamental plane.  
These constraints suggest that other forms of scatter (apart from bad sampling and missing flux) could be affecting the data. As summarised in KFC06 other forms of scatter could be attributed to beaming, source peculiarities and spectral energy distribution - but note, these should remain relatively constant for the $same$ source.

In figure \ref{KFC} we plot a LLAGN $only$ sample and best fit parameters taken from KFC06 with the updated BHXRB sample. 
Note, the X-ray data in the KFC06 sample are taken in the 1-10 keV band. As the NGC 7213 data points are in the 2-10 keV band we assume a photon index of $\Gamma$=1.8 and add a correction factor to the NGC 7213 data points to make them comparable. 
We apply a similar correction to the BHXRB sample.
From this plot it is clear that the NGC 7213 data points are positioned slightly above the best fit line.

Considering the postion of NGC 7213 on the fundamental plane with respect to other LLAGN (see figure 6), we calculate its radio loudness parameter to assess the differentiation. We calculate the radio loudness parameter $R = L_{6cm}/L_{B}$ (where $L_{6cm}$ and $L_{B}$ are the radio and optical luminosities); we use a $B$ band magnitude of 16.3, and find the optical flux $S_{opt}$ using $B = -2.5$log$S_{opt}-48.6$ \citep{Einstein_1}: giving $R$ = 134.8. In this scheme radio-loud sources are typically defined as having an $R$ parameter $>$ 10, while radio-quite range between 0.1-1 \citep{Kellermann}.  

Using the alternative radio loudness parameter of \cite{R_xray} which utilises the X-ray instead of optical luminosity, $R_{X} = L_{6cm}/L_{2-10 keV}$, we find log$R_{X} = $ -3.28. \cite{Panessa} show that for a sample of low-luminosity Seyfert Galaxies log$R_{X} = -3.64 \pm 0.16$ while for a sample of low-luminosity radio Galaxies (LLRGs) log$R_{X} = -1.40 \pm 0.11$. Therefore, with respect to the X-ray radio loudness, NGC 7213 is only slightly higher than that of a sample of low-luminosity Seyfert Galaxies; while under the standard definition of radio loudness it is indeed radio loud. These results are consistent with the position of NGC 7213 on figure \ref{KFC}.

\begin{figure}
\centering
\includegraphics[scale=0.88]{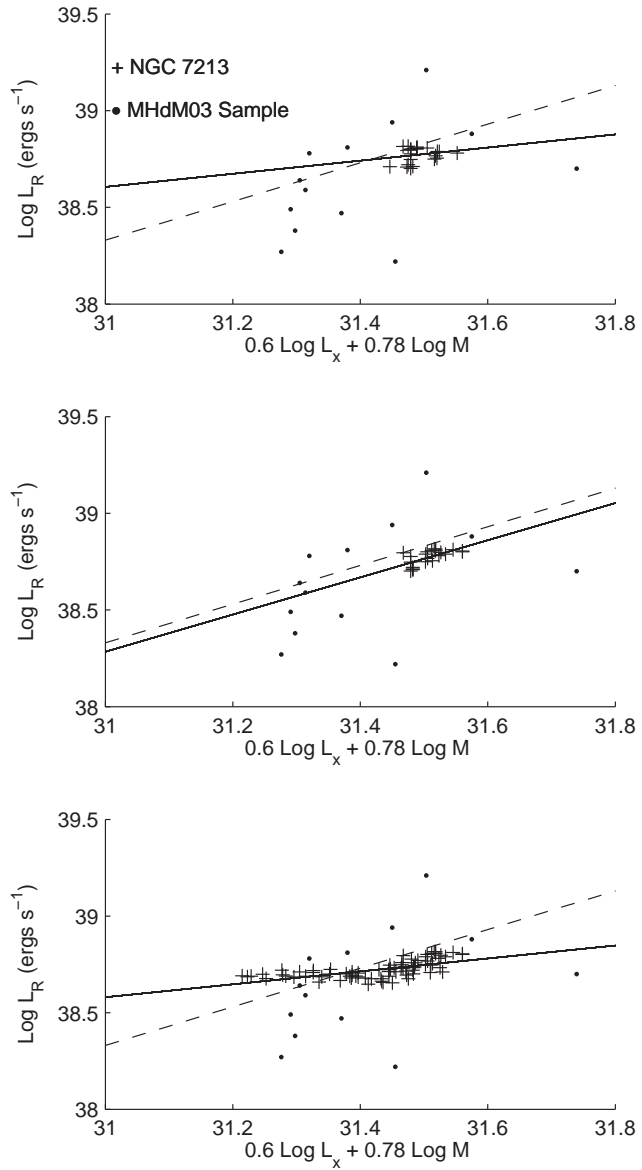}
\caption{Top Panel: Flare 1 radio data points at 4.8 GHz uncorrected for lag. On both graphs the solid line indicates the least squared best fit for the NGC 7213 data only; the dashed line indicates the MHdM03 best fit. Middle Panel: Flare 1 radio data points shifted -35 days to correct for time lag calculated from the cross-correlation function. Bottom Panel: All radio data points shifted back -35 days.}
\label{shifted}
\end{figure}

As was discussed earlier in this paper there is an apparent time lag between events in the X-ray and radio. 
Therefore comparing the fitting parameters found from the NGC 7213 data with the MHdM03 relationship without correcting for the lag might give rise to errors as we are not matching the correct data points. 
The width in the cross-correlation peaks shows that the time lag is variable. 
Thus, for example, the two radio flares could have different lag times associated with them.
Therefore shifting the entire radio light curve back by a set amount to match the X-ray could still give a scatter. 
To simplify this problem we separated out the data for the first flare only because we have a more accurate measurement of the lag in this specific case.
We then shifted the radio data -35 days which was the time lag measured for this singular flare using the DCF at 4.8 GHz (see table \ref{lag_table}).
The top panel of figure \ref{shifted} shows the uncorrected data on the MHdM03 plot while the middle panel shows the corrected data; for completeness the bottom panel shows all radio data points shifted back.

For the first flare correcting the data appears to reduce the scatter and increase the gradient more in line with the MHdM03 best fit. It is now described by $L_{R} = (0.58^{+0.14}_{-0.14})$ log $L_{X}+8.5^{+7.6}_{-7.6}$. 
To check the statistical significance of this we measured the gradient for a variety of shifts. From 0-25 days the gradient gradually steepens until it gets close to 1 (giving a coefficient of $\sim$ 0.6). From 30-50 days it plateaus $\sim$ 1 and from 50 days the gradient decreases towards 0.
Thus it does appear that moving the flare back by the amount given from the DCF function does seem to better represent the data with respect to the MHdM03 fit. However, it should be noted that as this is a log/log plot, measuring gradients from such a small range of values should be treated with care.
It would, however, be interesting in future studies to assess the importance of this parameter.

\section{Conclusion}
We have used the Australian Telescope Compact Array and the Rossi X-ray Timing Explorer to conduct a long term study of AGN variability in the LLAGN NGC 7213.
We have used the cross-correlation function to show that a complex and weakly significant correlated behaviour exists between the X-ray and radio emitting regions. 
Although the statistics only show a weak correlation, this study is the first definitive campaign to probe this type of behaviour. 

We have shown that NGC 7213 sits well on the predicted fundamental plane of black hole activity plot when compared with the MHdM03 sample.
However we have shown that when comparing NGC 7213 with a revised BHXRB and LLAGN sample that the data points are above the expected correlation; which is however consistent with the calculated radio and X-ray loudness parameters.
We have also shown some support that by correcting for the time lag between events in X-ray and radio the gradient of data points agree better with the best fit derived from the MHdM03 sample on the fundamental plane.

\section{Acknowledgements}
M.E.Bell would like to thank Sera Markoff, Anthony Rushton and Sadie Jones for their useful comments and discussion.
The Australia Telescope is funded by the Commonwealth of Australia for operation as a National Facility managed by CSIRO.
This research has made use of the Tartarus (Version 3.1) database, created by Paul O'Neill and Kirpal Nandra at Imperial College London, and Jane Turner at NASA/GSFC. Tartarus is supported by funding from PPARC, and NASA grants NAG5-7385 and NAG5-7067.

\appendix

\label{lastpage}


\begin{thebibliography}{99}


\bibitem[Alexander(1997)]{Alexander} Alexander, T.\ 1997, 
Astronomical Time Series, 218, 163 

\bibitem[Belloni(2010)]{Belloni_Review} Belloni, T.\ 2010, Lecture 
Notes in Physics, Berlin Springer Verlag, 794,  

\bibitem[\protect\citeauthoryear{Bianchi et 
al.}{2003}]{Bianchi_2003} Bianchi S., Matt G., Balestra I., Perola G.~C., 2003, A\&A, 407, L21 

\bibitem[\protect\citeauthoryear{Bianchi et 
al.}{2008}]{Bianchi_2008} Bianchi S., La Franca F., Matt G., 
Guainazzi M., Jimenez Bail{\'o}n E., Longinotti A.~L., Nicastro F., 
Pentericci L., 2008, MNRAS, 389, L52 

\bibitem[\protect\citeauthoryear{Bignall et 
al.}{2004}]{ATCA_Blazar} Bignall H.~E., Tzioumis A.~K., Jauncey 
D.~L., Venturi T., Clay R.~W., 2004, NuPhS, 132, 149 

\bibitem[\protect\citeauthoryear{Blandford 
\& Konigl}{1979}]{Bland_79} Blandford R.~D., Konigl A., 1979, ApJ, 232, 34 

\bibitem[\protect\citeauthoryear{Blank, Harnett, 
\& Jones}{2005}]{Blank_LBA} Blank D.~L., Harnett J.~I., Jones P.~A., 2005, MNRAS, 356, 734 

\bibitem[\protect\citeauthoryear{Bower, Falcke, 
\& Mellon}{2002}]{Percent_Flux_LLAGN_Bower} Bower G.~C., Falcke H., Mellon R.~R., 2002, ApJ, 578, L103 

\bibitem[\protect\citeauthoryear{Bransford et 
al.}{1998}]{Blanford_ACTA} Bransford M.~A., Appleton P.~N., Heisler 
C.~A., Norris R.~P., Marston A.~P., 1998, ApJ, 497, 133 

\bibitem[Breedt et al.(2009)]{Elme} Breedt, E., et al. 
2009, MNRAS, 394, 427 

\bibitem[Calvelo et al.(2010)]{Dan} Calvelo, D.~E., et al.\ 
2010, arXiv:1007.2313 

\bibitem[Casella et al.(2010)]{PG} Casella, P., et al.\ 
2010, arXiv:1002.1233 

\bibitem[\protect\citeauthoryear{Corbel et 
al.}{2003}]{Corbel_2003} Corbel S., Nowak M.~A., Fender R.~P., Tzioumis A.~K., Markoff S., 2003, A\&A, 400, 1007 

\bibitem[Dunn et al.(2009)]{Dunn} Dunn, R., Fender, R., 
Koerding, E., Belloni, T., \& Cabanac, C.\ 2009, arXiv:0912.0142 

\bibitem[Edelson 
\& Krolik(1988)]{CCF_Paper} Edelson, R.~A., Krolik, J.~H. 1988, ApJ, 333, 646 

\bibitem[\protect\citeauthoryear{Falcke, Patnaik, 
\& Sherwood}{1996}]{Falcke_1996} Falcke H., Patnaik A.~R., Sherwood W., 1996, ApJ, 473, L13 

\bibitem[\protect\citeauthoryear{Falcke, K{\"o}rding, 
\& Markoff}{2004}]{Falcke_2004} Falcke H., K{\"o}rding E., Markoff S., 2004, A\&A, 414, 895 

\bibitem[Falcke et 
al.(2009)]{SagA_Falcke} Falcke, H., Markoff, S., Bower, G.~C. 2009, AAP, 496, 77 

\bibitem[\protect\citeauthoryear{Felten}{1967}]{Felton} Felten 
J.~E., 1967, AJ, 72, 796 

\bibitem[Fender et al.(1999)]{Fender_1999} Fender, R.~P., 
Garrington, S.~T., McKay, D.~J., Muxlow, T.~W.~B., Pooley, G.~G., Spencer, 
R.~E., Stirling, A.~M., Waltman, E.~B. 1999, MNRAS, 304, 865 

\bibitem[Fender et al.(2007)]{Fender} Fender, R., Koerding, 
E., Belloni, T., Uttley, P., McHardy, I., 
\& Tzioumis, T.\ 2007, arXiv:0706.3838 

\bibitem[\protect\citeauthoryear{Ferrarese 
\& Merritt}{2000}]{Ferrarese} Ferrarese L., Merritt D., 2000, ApJ, 539, L9 

\bibitem[\protect\citeauthoryear{Gallo, Fender, 
\& Pooley}{2003}]{Gallo_2003} Gallo E., Fender R.~P., Pooley G.~G., 2003, MNRAS, 344, 60 

\bibitem[\protect\citeauthoryear{Gallo et al.}{2006}]{Gallo_2006} 
Gallo E., Fender R.~P., Miller-Jones J.~C.~A., Merloni A., Jonker P.~G., 
Heinz S., Maccarone T.~J., van der Klis M., 2006, MNRAS, 370, 1351 

\bibitem[Gallo(2007)]{Gallo_2007} Gallo, E.\ 2007, The 
Multicolored Landscape of Compact Objects and Their Explosive Origins, 924, 
715 

\bibitem[\protect\citeauthoryear{Halpern 
\& Filippenko}{1984}]{Einstein_1} Halpern J.~P., Filippenko A.~V., 1984, ApJ, 285, 475 

\bibitem[\protect\citeauthoryear{Hameed et al.}{2001}]{Liner_ref} 
Hameed S., Blank D.~L., Young L.~M., Devereux N., 2001, ApJ, 546, L97 

\bibitem[\protect\citeauthoryear{Jamil, Fender, 
\& Kaiser}{2009}]{Omar} Jamil O., Fender R., Kaiser C., 2009, arXiv, arXiv:0909.1309 

\bibitem[Jonker et al.(2010)]{Jonker} Jonker, P.~G., et al.\ 
2010, MNRAS, 401, 1255 

\bibitem[Kellermann et al.(1989)]{Kellermann} Kellermann, K.~I., 
Sramek, R., Schmidt, M., Shaffer, D.~B., \& Green, R.\ 1989, ApJ, 98, 1195

\bibitem[Koratkar 
\& Gaskell(1991)]{Koratkar} Koratkar, A.~P., Gaskell, C.~M. 1991, APJS, 75, 719 

\bibitem[\protect\citeauthoryear{K{\"o}rding, Falcke, 
\& Corbel}{2006}]{KFC} K{\"o}rding E., Falcke H., Corbel S., 2006, A\&A, 456, 439 

\bibitem[\protect\citeauthoryear{Maccarone, Gallo, 
\& Fender}{2003}]{Tom_Ed} Maccarone T.~J., Gallo E., Fender R., 2003, MNRAS, 345, L19 

\bibitem[Marscher et al.(2002)]{Marscher} Marscher, A.~P., 
Jorstad, S.~G., G{\'o}mez, J.-L., Aller, M.~F., Ter{\"a}sranta, H., Lister, 
M.~L., \& Stirling, A.~M. 2002, nat, 417, 625 

\bibitem[\protect\citeauthoryear{Markoff et 
al.}{2008}]{Markoff_M81} Markoff S., et al., 2008, ApJ, 681, 905 

\bibitem[\protect\citeauthoryear{McHardy et 
al.}{2006}]{Mc_Hardy_2006} McHardy I.~M., Koerding E., Knigge C., 
Uttley P., Fender R.~P., 2006, Natur, 444, 730 

\bibitem[\protect\citeauthoryear{Merloni, Heinz, 
\& di Matteo}{2003}]{Merloni} Merloni A., Heinz S., di Matteo T., 2003, MNRAS, 345, 1057 

\bibitem[Mirabel et 
al.(1998)]{Mirabel} Mirabel, I.~F., Dhawan, V., Chaty, S., Rodriguez, L.~F., Marti, J., Robinson, C.~R., Swank, J., \& Geballe, T. 1998, A\&P, 330, L9 

\bibitem[\protect\citeauthoryear{Nandra 
\& Pounds}{1994}]{Nandra} Nandra K., Pounds K.~A., 1994, MNRAS, 268, 405 

\bibitem[\protect\citeauthoryear{Nelson 
\& Whittle}{1995}]{Nelson} Nelson C.~H., Whittle M., 1995, ApJS, 99, 67

\bibitem[Panessa et 
al.(2007)]{Panessa} Panessa, F., Barcons, X., Bassani, L., Cappi, M., Carrera, F.~J., Ho, L.~C., \& Pellegrini, S. 2007, A\&A, 467, 519 

\bibitem[Peterson et al.(1998)]{Peterson} Peterson, B.~M., 
Wanders, I., Horne, K., Collier, S., Alexander, T., Kaspi, S., 
 Maoz, D. 1998, PASP, 110, 660 

\bibitem[Pooley 
\& Fender(1997)]{Pooley_Fender} Pooley, G.~G., Fender, R.~P. 1997, MNRAS, 292, 925 

\bibitem[Ptak(2001)]{LLAGN_Def} Ptak, A.\ 2001, X-ray Astronomy: 
Stellar Endpoints, AGN, and the Diffuse X-ray Background, 599, 326 

\bibitem[\protect\citeauthoryear{Rees}{1978}]{Rees} Rees 
M.~J., 1978, MNRAS, 184, 61P 

\bibitem[Russell et al.(2009)]{Dave} Russell, D.~M., Lewis, 
F., Roche, P., Clark, J.~S., Breedt, E., 
\& Fender, R.~P.\ 2009, arXiv:0911.4501 

\bibitem[\protect\citeauthoryear{Sault, Teuben, 
\& Wright}{1995}]{Sault} Sault R.~J., Teuben P.~J., Wright M.~C.~H., 1995, ASPC, 77, 433

\bibitem[Slee et al.(1994)]{Slee} Slee, O.~B., Sadler, 
E.~M., Reynolds, J.~E.,  Ekers, R.~D. 1994, MNRAS, 269, 928 

\bibitem[\protect\citeauthoryear{Spada et al.}{2001}]{Spada} 
Spada M., Ghisellini G., Lazzati D., Celotti A., 2001, MNRAS, 325, 1559 

\bibitem[\protect\citeauthoryear{Starling et 
al.}{2005}]{Starling} Starling R.~L.~C., Page M.~J., Branduardi-Raymont G., Breeveld A.~A., Soria R., Wu K., 2005, Ap\&SS, 300, 81 

\bibitem[\protect\citeauthoryear{Timmer 
\& Koenig}{1995}]{Timmer} Timmer J., Koenig M., 1995, A\&A, 300, 707 

\bibitem[Terashima 
\& Wilson(2003)]{R_xray} Terashima, Y., \& Wilson, A.~S.\ 2003, Active Galactic Nuclei: From Central Engine to Host Galaxy, 290, 403 

\bibitem[Thean et al.(2001)]{Thean_High_res} Thean, A., Pedlar, A., 
Kukula, M.~J., Baum, S.~A.,  O'Dea, C.~P.\ 2001, MNRAS, 325, 737 

\bibitem[\protect\citeauthoryear{Turner 
\& Pounds}{1989}]{Turner} Turner T.~J., Pounds K.~A., 1989, MNRAS, 240, 833

\bibitem[Turner et al.(2001)]{Tartarus} Turner, T.~J., Nandra, 
K., Turcan, D., 
\& George, I.~M.\ 2001, X-ray Astronomy: Stellar Endpoints, AGN, and the Diffuse X-ray Background, 599, 991 

\bibitem[Tzioumis(1997)]{LBA} Tzioumis, A.~K.\ 1997, Vistas 
in Astronomy, 41, 311 

\bibitem[van der Laan(1966)]{vanderlaan} van der Laan, H. 1966, 
NAT, 211, 1131 

\bibitem[Xue \& Cui(2007)]{Xue_XRB} Xue, Y.~Q., \& Cui, W. 2007, AAp, 466, 1053 

\end{thebibliography}
\end{document}